\documentclass[a4paper,12pt]{article}

\title{Near horizon superconformal symmetry of rotating\\ BPS black holes in five dimensions}
\author{Masayoshi Nakamura\footnote{E-mail: \tt mnakamura@hep1.c.u-tokyo.ac.jp}~ and~ 
Naoto Yokoi\footnote{E-mail: \tt nyokoi@tuhep.phys.tohoku.ac.jp}\bigskip\\
$^{*}$\textit{\small Institute of Physics, University of Tokyo, Tokyo 153-8902, Japan}\smallskip\\
$^{\dagger}$\textit{\small Department of Physics, Tohoku University, Sendai 980-8578, Japan}\smallskip\\
}

\usepackage{amsmath,amssymb,mathrsfs}
\usepackage{wrapfig}
\usepackage[dvips]{graphicx}
\usepackage{bm}

\begin{document}

\date{}

\maketitle

\begin{abstract}
We investigate 
the asymptotic supersymmetry group of the near horizon region of the BMPV black holes, 
which are the rotating BPS black holes in five dimensions.
When considering only bosonic fluctuations, 
we show that 
there exist consistent boundary conditions 
and the corresponding asymptotic symmetry group is generated by 
a chiral Virasoro algebra with the vanishing central charge.
After turning on fermionic fluctuations 
with the boundary conditions,
we also show that 
the asymptotic supersymmetry group is generated by 
a chiral super-Virasoro algebra with the vanishing central extension.
The super-Virasoro algebra is originated in the AdS$_2$ isometry supergroup 
of the near horizon solution.
\end{abstract}

\vspace{4cm}
\begin{flushright}
UT-Komaba 11-7\\
TU-891
\end{flushright}

\newpage

\section{Introduction}
BPS black holes in supergravity, which preserve a part of supersymmetry, 
play important roles in the understanding of the quantum mechanical nature of black holes. 
In the pioneering work by Strominger and Vafa \cite{Strominger:1996sh}, 
the five-dimensional BPS black hole is identified 
with a D-brane bound state in type IIB superstring theory and
the Bekenstein-Hawking entropy is explained by 
the microscopic counting in D-brane effective theory. 
The BPS black holes are also extensively studied 
from the perspective of the AdS/CFT correspondence \cite{Maldacena:1997re}.
In these analyses, supersymmetry is one of the keys to understanding 
the quantum properties of black holes in superstring theory. 

Recently, the Kerr/CFT correspondence, 
which is the duality 
between quantum gravity on the extremal Kerr black hole and 
a two-dimensional conformal field theory,  
has been suggested 
\cite{Guica:2008mu} 
(see \cite{Bredberg:2011hp} for a recent review).
This conjecture is based on the investigation of the asymptotic symmetry group 
on the near horizon geometry of the extremal Kerr black hole. 
(See section 3 for the definition of asymptotic symmetry group.)   
The asymptotic symmetry group is generated by a chiral Virasoro algebra with a nontrivial central extension, 
which gives a strong evidence for the Kerr/CFT correspondence.

The characteristic of the Kerr/CFT correspondence is that 
it does not require the BPS nature and the origin in D-branes of the black holes. 
If the Kerr/CFT correspondence is realized in supergravity (or superstring),
more information of quantum properties of black holes can be extracted, as in the AdS/CFT correspondence.
Supersymmetry will also play a key role in such a realization.  

However, in four dimensions, 
there is a theorem that 
asymptotically-flat rotating black holes cannot be supersymmetric,
i.e. cannot have any globally defined Killing spinors \cite{Gauntlett:1998fz}.
On the other hand, in five dimensions, there exists the asymptotically-flat 
BPS rotating black hole, so-called BMPV black hole \cite{Breckenridge:1996is}.
The BMPV black hole has been also investigated in the context of the Kerr/CFT correspondence 
\cite{Chow:2008dp,Isono:2008kx,Azeyanagi:2008dk,Chen:2009xja}.
Since the BMPV black hole is supersymmetric solution of supergravity, 
it is naturally expected that the asymptotic symmetry group is enhanced to 
a two-dimensional superconformal group, which is generated by a 
super-Virasoro algebra.\footnote{In three-dimensional AdS supergravity, the asymptotic symmetry group 
is known to be enhanced to a two-dimensional superconformal group, 
using the Chern-Simons formalism \cite{Banados:1998pi}.} 
So far, however, the {\it asymptotic supersymmetry group} has not been discussed 
from the perspective of the Kerr/CFT correspondence.\footnote{Asymptotic supersymmetry has been discussed in 
four-dimensional AdS space \cite{Henneaux:1985tv, Hollands:2006zu}. 
The resulting asymptotic supersymmetry group becomes an isometry supergroup of ${\rm AdS}_{4}$.}
Thus, in this paper, we discuss the BMPV black holes 
in five-dimensional minimal supergravity, focusing on the asymptotic supersymmetry group. 

Concretely, in the near horizon region of the BMPV black hole, 
we obtain the asymptotic Killing vectors under 
the specified boundary conditions for the metric and gauge field.
Conserved charges associated with the asymptotic symmetry group 
is constructed based on the covariant phase space method 
\cite{Lee:1990nz,Wald:1993nt,Iyer:1994ys}. 
The resulting charges satisfy the Virasoro algebra with vanishing central charge, 
which originates from the different boundary conditions from the Kerr/CFT case.
Furthermore, we obtain the asymptotic Killing spinors, which are related to the asymptotic supersymmetry group, 
under a boundary condition for gravitino. 
Applying the covariant phase space method to fermionic charges \cite{Hollands:2006zu}, 
we construct the conserved charges associated with the asymptotic Killing spinors 
and obtain the super-Virasoro algebra generating the asymptotic supersymmetry group.  

The organization of this paper is as follows.
In section \ref{sec:BMPV}, 
we summarize the basic properties of the BMPV black hole and its near horizon limit.
In section \ref{sec:ASG}, 
we study the asymptotic symmetry group of the near horizon solution, 
focusing on the bosonic fields. 
The effects of fermionic fields are considered in section \ref{sec:ASSG}.
The relation to other approaches to the BMPV black hole and 
the extension to other black holes are discussed in section \ref{sec:summary and discussion}. 
Technical tools are prepared in the appendices. 
In appendix \ref{sec:LL derivative}, 
an extension of Lie-derivative is introduced. 
The covariant phase space method is reviewed in appendix \ref{sec:conserved charges}. 
\\
\\
Our conventions are as follows:
\begin{itemize}
\item
We take the signature of the metric as $(-++++)$. 
We denote the local Lorentz indices as $a,b,c\cdots$ and 
the curved space indices as $\mu,\nu,\rho\cdots$.

\item
In this paper, 
we consider the torsion free situation exclusively.
In this situation, 
the spin connection is given by
\begin{align}
&\omega_{\mu a b}
=-\tfrac{1}{2}{e^{\nu}}_{a}(e_{b \mu,\nu}-e_{b \nu,\mu})
 -\tfrac{1}{2}{e^{\nu}}_{b}(e_{a \nu,\mu}-e_{a \mu,\nu})
 -\tfrac{1}{2}{e^{\rho}}_{a}{e^{\sigma}}_{b}(e_{d \sigma,\rho}-e_{d \rho,\sigma}){e^{d}}_{\mu},
\label{eq:spin}
\end{align}
where ${e^{a}}_{\mu}$ denotes the vielbein.

\item
The Clifford algebra is defined by 
$\{\Gamma^a,\Gamma^b\}=2\eta^{ab}$.
$\Gamma^{a_1\cdots a_n}$ denotes the completely antisymmetrized product, 
i.e. $\Gamma^{a_1\cdots a_n} \equiv \Gamma^{[a_1}\cdots\Gamma^{a_n]}$.
The hermiticity property is given by 
$(\Gamma^a)^{\dagger}=-\Gamma^0\Gamma^a (\Gamma^0)^{-1}$, 
and the Dirac conjugation is defined by 
$\overline{\psi}\equiv \psi^{\dagger}\Gamma^0$.
In our investigation, 
it is convenient to decompose a Dirac spinor $\psi$ as 
$\psi^{\pm}\equiv  \tfrac{i}{2}(1\pm i\Gamma^0)\psi$.

\item
We consider the various symmetry transformations in this paper.
For convenience, 
we distinguish them by the transformation parameters.
$v$ denotes the general coordinate transformation parameter; 
$\Lambda$ denotes the $U(1)$ gauge transformation parameter; and 
$\xi$ denotes the supersymmetry transformation parameter.
For example, 
$\delta_{v}g_{\mu\nu}$ means 
the general coordinate transformation of the metric.
\end{itemize}

\section{The BMPV solution}
\label{sec:BMPV}
In this section 
we review the BMPV black hole solution \cite{Breckenridge:1996is,Gauntlett:1998fz}.
It is the rotating BPS solution in $D=5$ minimal supergravity 
\cite{Cremmer:1980gs,Chamseddine:1980sp} described by the action
\begin{align}
S  
&=\frac{1}{16 \pi} \int d^5x \left[
  eR
 -eF_{\mu\nu}F^{\mu\nu}
 -2ie(\overline{\psi}_{\mu}\Gamma^{\mu\nu\rho}D_{\nu}\psi_{\rho}
     +\overline{\psi}_{\rho}\overleftarrow{D}_{\nu}\Gamma^{\mu\nu\rho}\psi_{\mu}) 
 +\sqrt{3}e\overline{\psi}_{\mu}X^{\mu\nu\rho\sigma}\psi_{\nu}F_{\rho\sigma}  
  \right] \notag  \\ 
&+\frac{1}{6\sqrt{3}\pi}\int A\wedge F\wedge F
 +\mathcal{O}(\psi_{\mu}^4),
\label{eq:action}
\end{align}
where 
$X^{\mu\nu\rho\sigma}
\equiv \Gamma^{\mu\nu\rho\sigma}+g^{\mu\rho}g^{\nu\sigma}-g^{\mu\sigma}g^{\nu\rho}$, 
and $D_{\mu}$ denotes the covariant derivative only containing the spin connection.
The supersymmetry transformation laws of this theory are given by 
\begin{align}
&\delta_{\xi} {e^{a}}_{\mu}
=i(\overline{\xi}\Gamma^{a}\psi_{\mu}-\overline{\psi}_{\mu}\Gamma^{a}\xi),
\label{eq:susy e}   \\
&\delta_{\xi} \psi_{\mu}
=D_{\mu}\xi
 +\tfrac{i}{4\sqrt{3}}({e^{a}}_{\mu}{\Gamma^{bc}}_{a}F_{bc}
           -4{e^{a}}_{\mu}\Gamma^{b}F_{ab})\xi
 +\mathcal{O}(\psi_{\mu}^2),
\label{eq:susy psi}   \\
&\delta_{\xi} A_{\mu}
=-\tfrac{\sqrt{3}}{2}(\overline{\psi}_{\mu}\xi-\overline{\xi}\psi_{\mu}).
\label{eq:susy A}
\end{align}
For our purpose 
the explicit forms of the higher order terms of the gravitinos are not important.

The BMPV solution is characterized by the two parameters $(\mu,j)$, 
which are related to the mass and the angular momentum. 
It is given by\footnotemark
\footnotetext{
Interesting geometric properties and causal structures of the BMPV black holes have been discussed in 
\cite{Gibbons:1999uv,Herdeiro:2000ap,Herdeiro:2002ft}.}
\begin{align}
ds^2
&=-\left(1-\frac{\mu}{r^2}\right)^2 \left(dt+\frac{j}{2(r^2-\mu)}\sigma_3\right)^2
  +\left(1-\frac{\mu}{r^2}\right)^{-2}dr^2
  +r^2d\Omega_3^2, 
\label{eq:BMPV g}\\
A
&=\frac{\sqrt{3}}{2}\left[
    \left(1-\frac{\mu}{r^2}\right)dt
   +\frac{j}{2r^2}\sigma_3 \right],  \qquad
\psi_{\mu}=0,
\label{eq:BMPV A psi}
\end{align}
where $d\Omega_3^2$ is the $3$-sphere metric and
$\sigma_3$ is one of the left invariant $1$-forms $\sigma_I$ ($I=1,2,3$).
It is convenient to parameterize the $3$-sphere by the Euler angles 
$(\theta,\phi,\psi)$ whose ranges are
\begin{align}
0 \leq \theta \leq \pi, \qquad 
0 \leq \phi \leq 2\pi, \qquad
0 \leq \psi \leq 4\pi.
\end{align} 
The left invariant $1$-forms are represented by
\begin{align}
&\sigma_1=-\sin{\psi}d\theta+\cos{\psi}\sin{\theta}d\phi,  \\
&\sigma_2= \cos{\psi}d\theta+\sin{\psi}\sin{\theta}d\phi,  \\
&\sigma_3=d\psi+\cos{\theta}d\phi,
\end{align}
and the $3$-sphere metric is given by
\begin{align}
d\Omega_3^2=\tfrac{1}{4}(d\theta^2+d\phi^2+d\psi^2+2\cos{\theta}d\psi d\phi).
\end{align}

The BMPV solution is supersymmetric 
because it has the Killing spinor 
which is the nontrivial solution of the Killing spinor equation
\begin{align}
D_{\mu}\xi
 +\tfrac{i}{4\sqrt{3}}({e^{a}}_{\mu}{\Gamma^{bc}}_{a}F_{bc}
           -4{e^{a}}_{\mu}\Gamma^{b}F_{ab})\xi
=0.
\label{eq:KSE}
\end{align}
The explicit form of the Killing spinor is
\begin{align}
\xi=\left(1-\frac{\mu}{r^2}\right)^{1/2}\eta^+, 
\end{align}
where $\eta$ is a constant Dirac spinor.

\subsection{The near horizon solution}
\label{sec:near horizon limit}
The BMPV black hole has the horizon which is located at $r=\sqrt{\mu}$.
Let us consider the near horizon limit which is given by 
making the coordinate transformations
\begin{align}
r\rightarrow \sqrt{\mu}\left(1+\tfrac{\lambda }{2}r\right),  \quad
t\rightarrow \tfrac{\sqrt{\mu}}{2\lambda}t,
\end{align}
and taking the limit $\lambda \rightarrow 0$.
In this limit, the BMPV solution (\ref{eq:BMPV g},\ref{eq:BMPV A psi}) reduces to
\begin{align}
ds^2
=-\frac{\mu}{4}\left(rdt+\frac{j}{\sqrt{\mu}^3}\sigma_3\right)^2
 +\frac{\mu}{4}\frac{dr^2}{r^2}
 +\mu d\Omega_3^2,  \qquad
A   
=\frac{\sqrt{3\mu}}{4}rdt+\frac{\sqrt{3}j}{4\mu}\sigma_3,  \qquad
\psi_{\mu}
=0.
\label{eq:near horizon BMPV}
\end{align}
This solution has the $SL(2,\mathbb{R}) \times SU(2) \times U(1)$ isometry group 
which is generated by the Killing vectors
\begin{align}
u_1 &= \partial_{t},  \label{eq:SL2 1} \\
u_2 &= t\partial_t-r\partial_r, \label{eq:SL2 2} \\ 
u_3 &= \tfrac{1}{2}\left[\tfrac{1}{r^2}\left(1-\tfrac{j^2}{\mu^3}\right)
                            +t^2\right]\partial_{t}
                            -tr \partial_{r}
                            +\tfrac{j}{\sqrt{\mu}^3r}\partial_{\psi},
\label{eq:SL2 3}  \\
v^L_1 &=\sin{\phi}\partial_{\theta}
       +\cot{\theta}\cos{\phi}\partial_{\phi}
       -\csc{\theta}\cos{\phi}\partial_{\psi},  \label{eq:SU2 1} \\
v^L_2 &=\cos{\phi}\partial_{\theta}
       -\cot{\theta}\sin{\phi}\partial_{\phi}
       +\csc{\theta}\sin{\phi}\partial_{\psi},  \label{eq:SU2 2}  \\
v^L_3 &=\partial_{\phi},
\label{eq:SU2 3}  \\
v^R_3 &=\partial_{\psi},
\label{eq:U1}
\end{align}
where $u_{I}$, $v_{I}^{L}$ and $v_{3}^{R}$ 
are the generators of the $SL(2,\mathbb{R})$, $SU(2)$ and $U(1)$ isometry group, 
respectively.

To identify the isometry supergroup of the near horizon solution (\ref{eq:near horizon BMPV}), 
we follow the arguments of \cite{Gauntlett:1998fz,Gauntlett:1998kc}.\footnotemark
\footnotetext{
The isometry supergroup is also investigated by the geometrical method for coset spaces 
\cite{AlonsoAlberca:2002wr,AlonsoAlberca:2002gh}.}
First, 
we need to find the Killing spinors 
on the near horizon solution (\ref{eq:near horizon BMPV}).
We choose the vielbeins as 
\begin{align}
e^0
=\tfrac{\sqrt{\mu}}{2}rdt+\tfrac{j}{2\mu}\sigma_3, \quad
e^1
=\tfrac{\sqrt{\mu}}{2}\sigma_1, \quad
e^2
=\tfrac{\sqrt{\mu}}{2}\sigma_2, \quad
e^3
=\tfrac{\sqrt{\mu}}{2}\sigma_3, \quad
e^r
=\tfrac{\sqrt{\mu}}{2r}dr.
\label{eq:bein}
\end{align}
and the product of all five gamma matrices as $\Gamma^{0123r}=i$.
Then the Killing spinor equation (\ref{eq:KSE}) reduces to
\begin{align}
&0=\left[\partial_{t}
           -\tfrac{r}{2}\Gamma^{0r}
           +\tfrac{ir}{2}\Gamma^{r}\right]\xi,  
\label{eq:reduced KSE t} \\
&0=\left[\partial_{r}
           -\tfrac{j}{2\sqrt{\mu}^3r}\Gamma^{03}
           -\tfrac{i}{2r}\Gamma^{0}
           +\tfrac{ij}{2\sqrt{\mu}^3r}\Gamma^{3}\right]\xi,  
\label{eq:reduced KSE r}  \\
&0=\left[\partial_{\theta}
           -\tfrac{\sin{\psi}}{2}M_1
           +\tfrac{\cos{\psi}}{2}M_2\right]\xi,  
\label{eq:reduced KSE theta}  \\
&0=\left[\partial_{\phi}
           -\tfrac{\cos{\theta}}{2}\Gamma^{21}
           +\tfrac{\sin{\theta}\cos{\psi}}{2}M_1
           +\tfrac{\sin{\theta}\sin{\psi}}{2}M_2\right]\xi,
\label{eq:reduced KSE phi}    \\
&0=\left[\partial_{\psi}
           -\tfrac{1}{2}\Gamma^{21}\right]\xi,
\label{eq:reduced KSE psi}
\end{align}
where 
\begin{align}
M_1 \equiv -\Gamma^{32}+\tfrac{j}{\sqrt{\mu}^3}\Gamma^{02}-\tfrac{ij}{\sqrt{\mu}^3}\Gamma^{2},
\qquad
M_2 \equiv \Gamma^{31}-\tfrac{j}{\sqrt{\mu}^3}\Gamma^{01}+\tfrac{ij}{\sqrt{\mu}^3}\Gamma^{1}.
\end{align}
The most general solution of eqs.(\ref{eq:reduced KSE t}-\ref{eq:reduced KSE psi}) is given 
by the linear combinations of the following two Killing spinors:
\begin{align}
\xi_1
&=r^{1/2}\Omega \eta^+, 
\label{eq:KS1} \\
\xi_2
&=\left[
  r^{-1/2}\left(-i+\tfrac{j}{\sqrt{\mu}^3}\Gamma^3\right)
 -t r^{1/2}\Gamma^r \right]\Omega \eta^-,
\label{eq:KS2}
\end{align}
where 
\begin{align}
\Omega
=e^{\frac{1}{2}\Gamma^{21}\psi}
 e^{\frac{1}{2}\Gamma^{13}\theta}
 e^{\frac{1}{2}\Gamma^{21}\phi},
\end{align}
and $\eta$ is an arbitrary constant Dirac spinor.
Next, 
we should consider the quantity
\begin{align}
\overline{\xi}\Gamma^{\mu}\xi'\partial_{\mu},
\label{eq:bilinear}
\end{align} 
where $\xi$ and $\xi'$ are Killing spinors.
As is discussed in \cite{Gauntlett:1998fz,Gauntlett:1998kc}, 
this is the Killing vector field
and generates the bosonic isometry group which is extended to the isometry supergroup.
In our case, 
the vector (\ref{eq:bilinear}) is spanned by 
the Killing vectors (\ref{eq:SL2 1}-\ref{eq:SU2 3}) only.
This implies that 
$SL(2,\mathbb{R})$ and $SU(2)$ enlarge to $SU(1,1|2)$,  
but $U(1)$ part remains the pure bosonic.
Therefore we can conclude that 
the isometry supergroup of the near horizon solution (\ref{eq:near horizon BMPV}) is
$SU(1,1|2)\times U(1)$.

\section{Asymptotic symmetry group}
\label{sec:ASG}
In this section 
we study fluctuations of the bosonic fields 
around the near horizon solution (\ref{eq:near horizon BMPV}).
In particular, we analyze the asymptotic symmetry group (ASG) in detail.
The ASG is defined by
the set of allowed symmetry transformations modulo 
the set of trivial symmetry transformations.
A transformation is allowed if it generates a fluctuation which obeys the boundary conditions.
A transformation is trivial if a conserved charge associated with it, 
which is defined in appendix \ref{sec:conserved charge and Poisson bracket}, 
vanishes.

In what follows 
we rename the near horizon fields given by eq.(\ref{eq:near horizon BMPV}) 
as $\bar{g}_{\mu\nu}$, $\bar{A}_{\mu}$ and $\bar{\psi}_{\mu}$, 
to emphasize that they are the background, 
and denote fluctuations around them by $h_{\mu\nu}$, $a_{\mu}$ and $\pi_{\mu}$, 
respectively.
Note that 
the total field configurations are given by
\begin{align} 
g_{\mu\nu}^{\text{(tot)}}
=\bar{g}_{\mu\nu}+h_{\mu\nu}, \qquad
A_{\mu}^{\text{(tot)}}
=\bar{A}_{\mu}+\partial_{\mu}\lambda +a_{\mu},  \qquad 
\psi_{\mu}^{\text{(tot)}}
=\bar{\psi}_{\mu}+\pi_{\mu},
\end{align}
where $\lambda=\lambda(t,\theta,\phi,\psi)$ is an $r$-independent arbitrary function 
which fixes a gauge of the background gauge field.\footnotemark
\footnotetext{In this context, 
we should regard $\bar{A}_{\mu}+\partial_{\mu}\lambda$ as the background.
The function $\lambda$ plays an important role in section \ref{sec:ASSG}.}
Throughout this section, 
we set $\pi_{\mu}$ to zero to focus on the bosonic fluctuations.
The fermionic fluctuations are considered in section 4.

\subsection{Boundary conditions}
We choose the boundary conditions
\begin{align}
h_{\mu\nu} \sim \mathcal{O}
\begin{pmatrix}
r & 1/r   & 1     & 1     & 1     \\
  & 1/r^4 & 1/r^2 & 1/r^2 & 1/r^2 \\
  &       & 1/r   & 1/r   & 1/r   \\
  &       &       & 1/r   & 1/r   \\
  &       &       &       & 1/r   \\
\end{pmatrix},
\quad
a_{\mu} \sim \mathcal{O}
\begin{pmatrix}
1& 1/r^2 & 1/r & 1/r & 1/r \\
\end{pmatrix},
\label{eq:BC for b}
\end{align}
in the basis $(t,r,\theta,\phi,\psi)$.
These boundary conditions are invariant under transformations 
generated by the Killing vectors 
(\ref{eq:SL2 1}-\ref{eq:U1}).
The most general allowed transformation $(v,\Lambda)$ is derived by 
solving the equations 
$h_{\mu\nu} \sim \delta_{v}g_{\mu\nu}^{\text{(tot)}}$ and 
$a_{\mu} \sim (\delta_{v} +\delta_{\Lambda}) A_{\mu}^{\text{(tot)}}$, 
or more explicitly
\begin{align}
h_{\mu\nu}
\sim \mathcal{L}_{v}(\bar{g}_{\mu\nu}+h_{\mu\nu}),  \qquad
a_{\mu}
\sim \mathcal{L}_{v}(\bar{A}_{\mu}+\partial_{\mu}\lambda+a_{\mu})+\partial_{\mu}\Lambda,
\label{eq:AKE and AGE}
\end{align} 
where $\mathcal{L}$ denotes the standerd Lie derivative.
Then we obtain the general solution 
\begin{align}
&v
=f(t)\partial_t-\partial_{t}f(t)r\partial_r+\sum^{3}_{I=1}g^I(t)v_I^L+h(t)v_3^R+ v^{\text{(sub)}},
\label{eq:AKV} \\
&\Lambda
=-v[\lambda]+\alpha (t)+\Lambda^{\text{(sub)}},
\label{eq:AGP}
\end{align}
where $f(t)$, $g^{I}(t)$, $h(t)$ and $\alpha(t)$ are arbitrary smooth functions 
and $v^{\text{(sub)}}$ and $\Lambda^{\text{(sub)}}$ denote the subleading terms 
which are given by
\begin{align}
&v^{\text{(sub)}}
=\mathcal{O}(1/r^2)\partial_{t}+\mathcal{O}(1/r)\partial_{r}
+\mathcal{O}(1/r)\partial_{\theta}+\mathcal{O}(1/r)\partial_{\phi}
+\mathcal{O}(1/r)\partial_{\psi},  \\
&\Lambda^{\text{(sub)}}=\mathcal{O}(1/r),
\end{align}
respectively.
Notice that 
the allowed transformation (\ref{eq:AKV}) includes 
all of the Killing vectors (\ref{eq:SL2 1}-\ref{eq:U1}).

\subsection{Conserved charges}
The conserved charges are defined and constructed 
in appendix \ref{sec:conserved charges}, 
using the covariant phase space method 
\cite{Lee:1990nz,Wald:1993nt,Iyer:1994ys,Hollands:2006zu}.
In this formalism, 
infinitesimal charge differences between $(g,A)$ and $(g+h,A+a)$ are given by\footnotemark
\footnotetext{
The conserved charge in more general theories containing $D=5$ minimal supergravity are derived in 
\cite{Compere:2009dp}, 
based on the slightly different method formulated in 
\cite{Barnich:2001jy,Barnich:2007bf,Compere:2007az}.
}
\begin{align}
\delta H_{(v,\Lambda)}
=\int_{\partial C}  \bm{k}_{v,\Lambda}(g,h;A,a),
\end{align}
where
\begin{align}
\bm{k}_{v,\Lambda}(g,h;A,a)
= \bm{k}_{v}^{E}(g,h)
 +\bm{k}_{v,\Lambda}^{F} (g,h;A,a)
 +\bm{k}_{v,\Lambda}^{CS}(A,a),
\end{align}
and 
\begin{align}
&\bm{k}_{v}^{E}(g,h)
 =\frac{\sqrt{-g}}{8\pi}\left[
  v^{\nu}\nabla^{\mu}h
 -v^{\nu}\nabla_{\rho}h^{\mu\rho}
 +v_{\sigma}\nabla^{\nu}h^{\mu\sigma}
 +\frac{1}{2}h\nabla^{\nu}v^{\mu}
 -h^{\nu\rho}\nabla_{\rho}v^{\mu}\right]
 \left(d^{3} x\right)_{\mu\nu},   \\
&\bm{k}_{v,\Lambda}^{F} (g,h;A,a)
 =\frac{\sqrt{-g}}{16\pi}\left[
  \left(-2hF^{\mu\nu}+8{h_{\rho}}^{\mu}F^{\rho\nu}-8\nabla^{\mu}a^{\nu}\right)
  (A_{\rho}v^{\rho}+\Lambda)  \right.  \notag \\
& \left.\hspace{8cm} 
 -4F^{\mu\nu}a_{\rho}v^{\rho}
 -8F^{\nu\rho}a_{\rho}v^{\mu}\right]
 \left(d^{3} x\right)_{\mu\nu},   \\
&\bm{k}_{v,\Lambda}^{CS}(A,a)
 =\frac{1}{\sqrt{3}\pi} a \wedge F(v\cdot A +\Lambda)
  +\frac{1}{3\sqrt{3} \pi}A\wedge a \wedge  \delta _{(v,\Lambda)}A,
\end{align}
where $\nabla_{\mu}$ denotes the covariant derivative only containing 
the Christoffel symbol. 
Covariant derivatives and raising or lowering indices are calculated by 
using $g_{\mu\nu}$.

The ASG is represented by the Poisson bracket algebra of the conserved charges, 
which is defined by
\begin{align}
\left[H_{(v,\Lambda)},H_{(v',\Lambda')}\right]_{PB}
\equiv \delta_{(v,\Lambda)} H_{(v',\Lambda')}.
\end{align}
As is explained in appendix \ref{sec:conserved charge and Poisson bracket}, 
this can be rewritten as\footnotemark
\footnotetext{
Here we assume that fluctuations on 
the near horizon solution (\ref{eq:near horizon BMPV}) satisfy 
the consistency condition (\ref{eq:consistency 3}), 
which is essential to make conserved charges and Poisson brackets well-defined.
See appendix \ref{sec:conserved charge and Poisson bracket} for details.}
\begin{align}
\left[H_{(v,\Lambda)},H_{(v',\Lambda')}\right]_{PB}
= H_{(v'',\Lambda'')}
 +K_{(v,\Lambda),(v',\Lambda')},
\label{eq:PBA BB}
\end{align}
where the central extension term $K_{(v,\Lambda),(v',\Lambda')}$ is given by
\begin{align}
K_{(v,\Lambda),(v',\Lambda')}
=\int_{\partial C} 
 \bm{k}_{v',\Lambda'}(g,h;A,a)
 |_{(h_{\mu\nu},a_{\mu})
    =(\delta_{v}+\delta_{\Lambda})(g_{\mu\nu},A_{\mu})},
\end{align}
and $(v'',\Lambda'')$ satisfies
\begin{align}
\delta_{(v'',\Lambda'')}
=\left[\delta_{(v,\Lambda)},\delta_{(v',\Lambda')}\right],
\label{eq:b b commutator}
\end{align}
for the background configurations.
From the direct computation, 
we have 
\begin{align}
v''
=\left[v,v'\right]_{LB},  \qquad
\Lambda''
=v[\Lambda']-v'[\Lambda], 
\label{eq:Lie bracket}
\end{align}
where 
$\left[\cdot,\cdot\right]_{LB}$ denotes the Lie bracket.

By calculating the conserved charges associated with the allowed transformations 
(\ref{eq:AKV},\ref{eq:AGP}), 
we find that they do not diverge and do not vanish only for 
\begin{align}
v=f(t)\partial_t-\partial_{t}f(t)r\partial_r,  \qquad
\Lambda=-v[\lambda].
\label{eq:ASG generator}
\end{align}
This means that the ASG is generated by the transformations (\ref{eq:ASG generator}).
To identify the ASG, 
it is convenient to expand $f(t)$ in terms of the Laurent series
\begin{align}
f(t)
=\sum_{m \in \mathbb{Z}} i t^{m+1}f_{m},
\end{align}
where $f_{m}$ are pure imaginary constants.
Then the conserved charge associated with the transformation (\ref{eq:ASG generator}) 
are written by
\begin{align}
H_{(v,\Lambda)}
=\sum_{m \in \mathbb{Z}} f_{m}
 H_{(v_{m},-v_{m}[\lambda])}
\label{eq:expanded Hv}
\end{align}
where $v_{m}$ are defined by
\begin{align}
v_{m}
=i t^{m+1}\partial_{t} -i(m+1)t^{m} r \partial_{r}.
\end{align}

For simplicity, 
we redefine $L_{m}=H_{(v_{m},-v_{m}[\lambda])}$.
Then, 
in the same way, 
$H_{(v',-\Lambda')}$ and $H_{(v'',\Lambda'')}$ 
are expanded as
\begin{align}
H_{(v',\Lambda')}
=\sum_{m \in \mathbb{Z}} f'_{m} L_{m},  \qquad
H_{(v'',\Lambda'')}
=\sum_{m,n \in \mathbb{Z}} f_{m}f'_{n} (-i)(m-n) L_{m+n},
\end{align}
respectively.
Noting that  
$K_{(v,\Lambda),(v',\Lambda')}$ vanishes for the transformation (\ref{eq:ASG generator}), 
we find
\begin{align}
\sum_{m,n \in \mathbb{Z}}  f_{m} f'_{n} 
\left[L_{m},L_{n} \right]_{PB}
=\sum_{m,n \in \mathbb{Z}} 
 f_{m}f'_{n} (-i)(m-n) L_{m+n},
\end{align}
or equivalently,
\begin{align}
i \left[
L_{m},L_{n} \right]_{PB}
= (m-n) L_{m+n}.
\label{eq:classical ASG}
\end{align}
By the semiclassical quantization procedure 
which consists of the replacement $[\cdot,\cdot]_{PB}\rightarrow \tfrac{1}{i}[\cdot,\cdot]$ 
and the reinterpretation of the conserved charges 
$L_{m}$ as the quantum operators $\hat{L}_{m}$, 
we have the quantum version
\begin{align}
[\hat{L}_m,\hat{L}_n ]
=(m-n) \hat{L}_{m+n}.
\label{eq:quantum ASG}
\end{align}
This is the chiral Virasoro algebra without the central extension.\footnotemark
\footnotetext{
The general solution (\ref{eq:AKV},\ref{eq:AGP}) also contains
\begin{align}
v^{KM}
=\sum^{3}_{I=1}g^I(t)v_I^L+h(t)v_3^R, \qquad
\Lambda^{KM}
=-v^{KM}[\lambda]+\alpha (t),
\notag
\end{align}
and $(v^{KM}, \Lambda^{KM})$ obey the $\widehat{su}(2)\times \widehat{u}(1) \times \widehat{u}(1)$ Kac-Moody algebra 
under the Lie bracket (\ref{eq:Lie bracket}).
However, 
these parameters only generate trivial transformations.}

\section{Asymptotic supersymmetry group}
\label{sec:ASSG}
Let us move on the study of fermionic fluctuations 
around the background (\ref{eq:near horizon BMPV}).
In this section 
we identify the asymptotic supersymmetry group (ASSG) 
which is defined in a similar way to the ASG.
In particular, 
we find the two-dimensional superconformal group, 
which is the supersymmetric extension of the ASG derived in section \ref{sec:ASG}.

\subsection{Boundary conditions}
In principle, 
to find the most general allowed transformation, 
we must solve the equations
\begin{align}
h_{\mu\nu}
&\sim \left(\delta_{v}+\delta_{\xi}\right)g_{\mu\nu}^{\text{(tot)}},  
\label{eq:full AKE} \\
a_{\mu}
&\sim \left(\delta_{v}+\delta_{\Lambda}+\delta_{\xi}\right)A_{\mu}^{\text{(tot)}}, 
\label{eq:full AGE} \\
\pi_{\mu}
&\sim \left(\delta_{v}+\delta_{\xi}\right)\psi_{\mu}^{\text{(tot)}},
\label{eq:full AKSE}
\end{align}
under appropriate boundary conditions.
Then we need to deal with the finite fluctuations of gravitinos in the bulk, 
but it is obvious that these fluctuations violate the torsion free condition.
To avoid this undesirable situation, 
we assume that fluctuations of all fields are infinitesimal everywhere.
Under this assumption, 
it is only necessary to analyze eqs.(\ref{eq:full AKE}-\ref{eq:full AKSE}) 
at the linearized level with respect to $(h_{\mu\nu},a_{\mu},\pi_{\mu})$ and $(v,\Lambda,\xi)$, 
and these eqations reduce to
\begin{align}
h_{\mu\nu}
\sim \mathcal{L}_{v}\bar{g}_{\mu\nu},  \qquad
a_{\mu}
\sim \mathcal{L}_{v}\left(\bar{A}_{\mu}+\partial_{\mu} \lambda\right)
     +\partial_{\mu}\Lambda,
\label{eq:reduced AKE and AGE}
\end{align}
and
\begin{align}
\pi_{\mu}
&\sim \left[
      \bar{D}_{\mu}
     +\tfrac{i}{4\sqrt{3}}({\bar{e}^{a}}_{\mu}{\Gamma^{bc}}_{a}\bar{F}_{bc}
                         -4{\bar{e}^{a}}_{\mu}\Gamma^{b}\bar{F}_{ab})
      \right] \xi.
\label{eq:reduced AKSE}
\end{align}
Since eqs.(\ref{eq:reduced AKE and AGE}) have no fluctuations in the right-hand side, 
these equations are different from eqs.(\ref{eq:AKE and AGE}).
However, 
we can show that 
the general solution of eqs.(\ref{eq:reduced AKE and AGE}) is also given by 
eqs.(\ref{eq:AKV},\ref{eq:AGP}) under the same boundary conditions (\ref{eq:BC for b}).
This means that the results for the ASG derived in section \ref{sec:ASG} remain valid 
in the following analysis of the ASSG.
Therefore, 
in this section, 
we concentrate on analyzing the effects of supersymmetry transformations 
derived from eq.(\ref{eq:reduced AKSE}).

For fluctuations of the gravitinos, 
we choose the boundary conditions\footnotemark
\footnotetext{
The boundary conditions (\ref{eq:BC for f}) are invariant under transformations 
generated by the Killing vectors (\ref{eq:SL2 1}-\ref{eq:U1}).
However, in our linearized analysis, this property does not play an essential role. }
\begin{align}
&
\pi_{t}^{+}
\sim \mathcal{O}(r^{-1/2}), \quad
\pi_{r}^{+}
\sim \mathcal{O}(r^{-3/2}), \quad
\pi_{\theta}^{+}
\sim \mathcal{O}(r^{-1/2}), \quad
\pi_{\phi}^{+}
\sim \mathcal{O}(r^{-1/2}), \quad
\pi_{\psi}^{+}
\sim \mathcal{O}(r^{-1/2}), \notag \\
&
\pi_{t}^{-}
\sim \mathcal{O}(r^{-1/2}), \quad
\pi_{r}^{-}
\sim \mathcal{O}(r^{-5/2}), \quad
\pi_{\theta}^{-}
\sim \mathcal{O}(r^{-3/2}), \quad
\pi_{\phi}^{-}
\sim \mathcal{O}(r^{-3/2}), \quad
\pi_{\psi}^{-}
\sim \mathcal{O}(r^{-3/2}).
\label{eq:BC for f}
\end{align}
Under the above boundary conditions, 
the most general solution of eq.(\ref{eq:reduced AKSE}) is given by 
\begin{align}
\xi=-i(\xi^++\xi^-)
\label{eq:full AKS}
\end{align} 
where
\begin{align}
\xi^{+}=r^{1/2}\Omega \eta^{+}(t) 
       +\mathcal{O}(r^{-1/2}),  \qquad
\xi^{-}=i r^{-1/2}\Omega \Gamma^{r} \partial_{t} \eta^{+}(t) 
       +\mathcal{O}(r^{-3/2}),
\label{eq:AKS}
\end{align}
and $\eta(t)$ is an arbitrary smooth Dirac spinor function.
Notice that 
the general solution (\ref{eq:full AKS}) includes 
all of the Killing spinors (\ref{eq:KS1},\ref{eq:KS2}).

\subsection{Conserved charges}
According to appendix \ref{sec:conserved charges},
infinitesimal charge differences 
between the $\psi_{\mu}=0$ background and fluctuated configurations $\pi_{\mu}$ 
are given by
\begin{align}
\delta H_{\xi}
=\int_{\partial C} \bm{k}_{\xi}(\pi),
\end{align}
where\footnotemark
\footnotetext{
Our choice (\ref{eq:bein}) makes $e$ negative, 
so the volume element should be given by $|e|$ rather than $e$.}
\begin{align}
\bm{k}_{\xi}(\pi)
\equiv \bm{k}^{\psi}_{\xi}(\psi,\pi)|_{\psi=0} 
     =-\tfrac{i}{4\pi} |e| \overline{\xi} \Gamma^{\mu\nu\rho} \pi_{\rho}
       \left(d^3 x\right)_{\mu\nu} 
      + \text{h.c.}
\end{align}
Along with the ASG, 
the ASSG is also generated by the Poisson bracket algebra of the conserved charges.
In this case we need to consider two types of Poisson brackets: 
$[H_{\xi},H_{\xi'}]_{PB}$ and $[H_{(v,\Lambda)},H_{\xi}]_{PB}$.
The former bracket is given by
\begin{align}
\left[H_{\xi},H_{\xi'}\right]_{PB}
=H_{(\tilde{v},\tilde{\Lambda})}+K_{\xi,\xi'},
\label{eq:PB xixi}
\end{align}
where the central extension term $K_{\xi,\xi'}$ is given by
\begin{align}
K_{\xi,\xi'}
=\int_{\partial C} \bm{k}_{\xi'}(\pi)|_{\pi_{\mu}=\delta_{\xi}\psi_{\mu}},
\end{align}
and $(\tilde{v},\tilde{\Lambda})$ satisfies 
\begin{align}
\delta_{(\tilde{v},\tilde{\Lambda})}=[\delta_{\xi},\delta_{\xi'}]
\label{eq:closure}
\end{align}
for the background configurations.
According to the closure relation \cite{Cremmer:1980gs} of $D=5$ minimal supergravity, 
the right-hand side of eq.(\ref{eq:closure}) is expanded by 
the general coordinate transformation, 
$U(1)$ gauge transformation, 
the supersymmetry transformation 
and the local Lorentz transformation.
Since the supersymmetry transformation parameter is given by $\mathcal{O}(\psi_{\mu}^1)$, 
it vanishes on the $\psi_{\mu}=0$ background.
Although the local Lorentz transformation parameter is given by $\mathcal{O}(\psi_{\mu}^0)$, 
the conserved charge associated with it vanishes on the $\psi_{\mu}=0$ background.
Thus we can neglect the latter two transformations 
and can read off $(\tilde{v},\tilde{\Lambda})$ 
by comparison with the closure relation as follows:
\begin{align}
\tilde{v}
=-\left(i\overline{\xi} \Gamma^{\mu} \xi' 
       -i\overline{\xi'} \Gamma^{\mu} \xi  \right) \partial_{\mu}, \qquad
\tilde{\Lambda}
=-\tfrac{\sqrt{3}}{2}\left(\overline{\xi}\xi' -\overline{\xi'}\xi \right)
 -A_{\mu} \tilde{v}^{\mu}.
\label{eq:v Lambda tilde}
\end{align}
The latter bracket is given by 
\begin{align}
\left[H_{(v,\Lambda)},H_{\xi}\right]_{PB}
=H_{\tilde{\xi}},
\label{eq:PB vxi}
\end{align}
where $\tilde{\xi}$ satisfies 
\begin{align}
\delta_{\tilde{\xi}}
=[\delta_{(v,\Lambda)},\delta_{\xi}],
\label{eq:b f commutator}
\end{align}
for the background configurations. 
Notice that 
it is difficult to derive $\tilde{\xi}$ from eq.(\ref{eq:b f commutator}) directly, 
because we do not know how the supersymmetry transformation acts on 
the symmetry transformation parameters.
However, as is discussed in \cite{Ortin:2002qb}, 
it seems reasonable that $\tilde{\xi}$ is given by
\begin{align}
\tilde{\xi}
=\mathbb{L}_{v}\xi,
\label{eq:Lv xi}
\end{align}
where $\mathbb{L}$ denotes the Lie-Lorentz derivative \cite{Ortin:2002qb} 
reviewed in appendix \ref{sec:LL derivative}.
We adopt this expression in this paper,  
even if $v$ is not only the Killing vector but also the asymptotic Killing vector generating the ASG.

By calculating the conserved charges associated with the allowed transformation (\ref{eq:AKS}), 
we find that they do not diverge 
and do not vanish only for 
\begin{align}
\xi^{+}=r^{1/2}\Omega \eta^{+}(t),  \qquad
\xi^{-}=i r^{-1/2}\Omega \Gamma^{r} \partial_{t} \eta^{+}(t).
\label{eq:ASSG generator}
\end{align}
This implies that these spinors generate the ASSG.
For the transformations (\ref{eq:ASG generator}) and (\ref{eq:ASSG generator}), 
eq.(\ref{eq:Lv xi}) reduces to
\begin{align}
\mathbb{L}_{v} \xi
=-i r^{1/2}\left[f(t) \Omega \partial_{t} \eta^+ (t) 
                -\tfrac{1}{2}\partial_{t}f(t)\Omega \eta^+ (t)\right],
\end{align}
up to trivial parts.
Furthermore, 
for the spinors (\ref{eq:ASSG generator}), 
eqs.(\ref{eq:v Lambda tilde}) reduce to
\begin{align}
\tilde{v}
&=\tilde{f}(t)\partial_{t}
 -\partial_{t}\tilde{f}(t)r\partial_{r},
\label{eq:v tilde} \\
\tilde{\Lambda}
&=-\tilde{v}[\lambda]
  -\tfrac{\sqrt{3}j}{4 \mu} 
   \left(-\tilde{g}^{1}(t)\cos{\phi}\sin{\theta}
         +\tilde{g}^{2}(t)\sin{\phi}\sin{\theta}
         +\tilde{g}^{3}(t)\cos{\theta}\right),
\label{eq:Lambda tilde}
\end{align}
where the trivial generators are neglected,
and 
\begin{align}
\tilde{f}(t)
&=\tfrac{2}{\sqrt{\mu}}
 \left(-\overline{\eta^+(t)}\eta'^+(t) +\text{h.c.}\right),  \\
\tilde{g}^{I}(t)
&=\tfrac{2}{\sqrt{\mu}}
 \left(\overline{\eta^+(t)}\Gamma^{I}\Gamma^{r}\partial_{t}\eta'^+(t)
      -\overline{\partial_{t}\eta^+(t)}\Gamma^{I}\Gamma^{r}\eta'^+(t) 
      +\text{h.c.}\right).
\end{align}
Eq.(\ref{eq:v tilde}) and the first term of eq.(\ref{eq:Lambda tilde}) 
can be interpreted as the ASG generators 
since these parts are of the same forms as eqs.(\ref{eq:ASG generator}).
On the other hand, 
the second term of eq.(\ref{eq:Lambda tilde}) should be viewed as 
the gauge transformation acting on the background gauge field.
Then the background gauge fixing parameter $\lambda$ is shifted.
Noting that the discussion of section \ref{sec:ASG} is applicable for any $r$-independent $\lambda$, 
it is clear that 
the second term of eq.(\ref{eq:Lambda tilde}) does not affect the discussion of the ASSG.
Therefore we can neglect this term in the following analysis.

To identify the ASSG, 
it is convenient to expand $\eta(t)$ as
\begin{align}
\eta(t)
=\sqrt{2} \mu^{1/4} 
 \sum_{p \in \mathbb{Z}+1/2} t^{p+1/2} \eta_{p},
\end{align}
where $\eta_{p}$ are constant Dirac spinors.
Now $\xi=-i(\xi^+ +\xi^-)$ reduces to
\begin{align}
\xi
=\sqrt{2} \mu^{1/4} 
 \sum_{p \in \mathbb{Z}+1/2} 
 \left(\xi_{p} + \xi^{sub}_{p} \right) \eta_{p}^+
\end{align}
where
\begin{align}
\xi_{p}
=-ir^{1/2}\Omega t^{p+1/2}, \qquad
\xi_{p}^{sub}
=(p+1/2) r^{-1/2}\Omega \Gamma^{r} t^{p-1/2}.
\end{align}
Furthermore $H_{\xi}$ is expanded as
\begin{align}
H_{\xi}
=\sum_{p \in \mathbb{Z}+1/2} \left(
 \overline{\eta_{p}^{+}}G_{p} 
-\overline{G_{p}} \eta_{p}^{+}\right),
\label{eq:expanded Hxi}
\end{align}
where
\begin{align}
G_{p}
=\sqrt{2} \mu^{1/4}  
 \int_{\partial C} -\frac{i}{4 \pi}|e|\xi_{p}^{\dagger} \Gamma^{\mu\nu\rho}\psi_{\rho}
 \left(d^3 x\right)_{\mu\nu}.
\end{align}
In a similar way, 
$H_{\mathbb{L}_{v}\xi}$ and $H_{\left(\tilde{v},-\tilde{v}[\lambda]\right)}$ are expanded as
\begin{align}
&H_{\mathbb{L}_{v}\xi}
 =\sum_{m \in \mathbb{Z}} \sum_{p\in \mathbb{Z}+1/2} 
  f_{m} \overline{\eta_{p}^{+}} \cdot i \left(p-\tfrac{m}{2}\right) G_{m+p} 
 +\text{h.c.},  
\label{eq:expanded HLvxi} \\
&H_{\left(\tilde{v},-\tilde{v}[\lambda]\right)}
 =4 \sum_{p,q \in \mathbb{Z}+1/2} 
  \left(-\overline{\eta_{p}^+} \eta_{q}'^+ +\overline{\eta_{q}'^+} \eta_{p}^+ \right)
  (-i)L_{p+q},
\label{eq:expanded Hxixi}
\end{align}
respectively.
Since the conserved charges 
(\ref{eq:expanded Hv},\ref{eq:expanded Hxi},\ref{eq:expanded HLvxi},\ref{eq:expanded Hxixi}) 
satisfy eqs.(\ref{eq:PB xixi},\ref{eq:PB vxi}), 
we could derive the Poisson bracket algebras analogous to eq.(\ref{eq:classical ASG}) 
by removing the expansion coefficients.
However, 
rather than doing this, 
we directly derive the quantum (anti)commutation relations anologous to 
eq.(\ref{eq:quantum ASG}).
To this end, 
in addition to the replacement implemented in section \ref{sec:ASG}, 
we replace $G_{p}$ and $\eta_{p}^{+}$ by the real Grassmann operators $\hat{G}_{p}$ 
and the real Grassmann numbers $\alpha_{p}$, respectively.\footnotemark
\footnotetext{
Although G$_p$ has four components, 
we will focus on the component 
which corresponds to a fermionic generator of the minimal extension of Virasoro algebra. 
Other components will not be essential in the following analysis.}
Furthermore, 
noting that the central extension term $K_{\xi,\xi'}$ vanishes for the transformations (\ref{eq:ASSG generator}), 
we have 
\begin{align}
 \sum_{m \in \mathbb{Z}} \sum_{p \in \mathbb{Z}+1/2} [f_{m} \hat{L}_{m},\alpha_{p} \hat{G}_{p}]
&=\sum_{m \in \mathbb{Z}} \sum_{p \in \mathbb{Z}+1/2} 
  f_{m} \left(\tfrac{m}{2}-p\right)\alpha_{p} \hat{G}_{m+p}, \\
 \sum_{p,q \in \mathbb{Z}+1/2}  
 [\alpha_{p} \hat{G}_{p}, \alpha'_{q} \hat{G}_{q}]
&=\sum_{p,q \in \mathbb{Z}+1/2} -\alpha_{p} \alpha'_{q} \cdot 2 \hat{L}_{p+q}.
\end{align}
By removing the paremeters $f_{m}$ and $\alpha_{p}$, these equations reduce to
\begin{align}
[\hat{L}_{m}, \hat{G}_{p}]
=\left(\tfrac{m}{2}-p\right) \hat{G}_{m+p}, \qquad
\{\hat{G}_{p},\hat{G}_{q}\}
=2 \hat{L}_{p+q}.
\label{eq:quantum ASSG}
\end{align}
Thus, 
in conjunction with eq.(\ref{eq:quantum ASG}), 
the quantum operators $\hat{L}_{m}$ and $\hat{G}_{p}$ satisfy 
the super-Virasoro algebra without the central extension.

\section{Summary and discussion}
\label{sec:summary and discussion}
We have investigated the asymptotic symmetry group 
in the near horizon region of the BPS rotating black hole in five dimensions.
After obtaining the asymptotic Killing vectors under 
the specified boundary conditions for the graviton and gauge field, 
we have constructed the finite and nonvanishing conserved charges associated with 
the asymptotic Killing vectors and found that the resulting charges obey 
the Virasoro algebra with the vanishing central extension.
Next, we obtained the asymptotic Killing spinors at the linearized level, 
under the suitable boundary conditions for the gravitino. 
Based on the covariant phase space method, we have also 
constructed the fermionic charges associated with the asymptotic Killing spinors
and found that these charges generate the super-Virasoro algebra 
with the vanishing central extension, together with the bosonic charges.
This asymptotic super-Virasoro algebra will shed some light on 
the quantum mechanics of the BMPV black holes.
 
Here are some discussions on the relation to other approaches to black holes.

\paragraph{Relation to Kerr/CFT correspondence}

The BMPV black hole in five dimensions has been 
investigated in the context of the Kerr/CFT correspondence 
\cite{Chow:2008dp, Isono:2008kx, Azeyanagi:2008dk, Chen:2009xja}.
In those analyses, an asymptotic Virasoro algebra with a 
non-vanishing central charge is obtained and 
the Bekenstein-Hawking entropy of the black hole is reproduced  
by the Cardy formula of the hypothetical dual conformal field theory in two dimensions. 
Although the asymptotic super-Virasoro algebra obtained in this paper includes
the Virasoro algebra, there are some crucial differences from the Kerr/CFT analysis.
One is the boundary condidtion (\ref{eq:BC for b}) for the metric and gauge field 
and the other is the geometric origin of the Virasoro algebra. 
The Virasoro algebra discussed in section \ref{sec:ASG} is associated with  
the asymptotic Killing vector in the time and radial direction  
and includes the isometry $SL(2,\mathbb{R})$ of the near horizon solution. 
On the other hand, the Virasoro algebra discussed in 
\cite{Chow:2008dp, Isono:2008kx, Azeyanagi:2008dk, Chen:2009xja} 
is associated with the asymptotic Killing vector in the angular direction
and completely decoupled from the $SL(2,\mathbb{R})$ isometry. 
Also, it is discussed the existence of two choices of the asymptotic 
Virasoro algebra whose zero mode associated with $\partial_{\phi}$ or $\partial_{\psi}$.  
From the perspective of supersymmetry, the Killing vector in 
the angular direction $\partial_{\psi}$ has nothing to do with the isometry supergroup 
(see section \ref{sec:BMPV}) and another Killing vector $\partial_{\phi}$ is a part of 
the R-symmetry of the isometry supergroup $SU(1,1|2)$.
We showed that our boundary condition (\ref{eq:BC for b}) allows the supersymmetric extension 
of the asymptotic Virasoro algebra based on the isometry $SL(2,\mathbb{R})$.      
It is very interesting to search for another boundary condition which 
allows an asymptotic supersymmetry based on the Killing vector $\partial_{\phi}$ 
which should have a nonvanishing central extension.

Another interesting problem is the extension of the analysis presented here 
to Kerr black holes in four dimensions.
Since our analysis focuses on the near horizon geometry of the Kerr black hole, 
which is no longer asymptotically-flat, it will be possible to find out 
the asymptotic Killing spinors and the associated super-Virasoro algebra. 

\paragraph{Relation to ${\rm AdS}_{2}/{\rm CFT}_{1}$ correspondence}

As is well-known, the near horizon geometry of the 
BMPV black hole has an ${\rm AdS}_{2}$ factor, whose isometry is $SL(2, \mathbb{R})$.
Since our asymptotic symmetry group includes this isometry $SL(2, \mathbb{R})$, 
one can naturally interpret our results from the ${\rm AdS}_{2}/{\rm CFT}_{1}$ correspondence 
\cite{Maldacena:1997re} (see also \cite{Strominger:1998yg, Nakatsu:1998st}), 
which is the duality between gravity on this near horizon geometry and 
a conformally invariant quantum mechanics (CQM).
In the context of the ${\rm AdS}_{2}/{\rm CFT}_{1}$ correspondence, 
the asymptotic analysis discussed in this paper implies 
that the dual CQM has an infinite dimensional super-Virasoro symmetry.
(Such a super-Virasoro algebra is discussed in supersymmetric CQM \cite{Marcus:2001rw}.)
If this is true, the super-Virasoro algebra will be very useful to obtain 
the spectrum and correlation functions of the dual CQM \cite{Chamon:2011xk, Chappell:2008in}.

Based on the ${\rm AdS}_{2}/{\rm CFT}_{1}$ correspondence, 
various approaches to the microscopic origin of Bekenstein-Hawking entropy are discussed;
Approaches based on the quantum entropy function \cite{Sen:2008yk, Sen:2009vz, Banerjee:2009af} 
and based on the entanglement entropy \cite{Azeyanagi:2007bj}, 
and the probe D0-brane approach \cite{Gaiotto:2004pc, Gaiotto:2004ij}.
It is very interesting to understand the relationship between 
our analysis in this paper and these approaches. 
In particular, these approaches have been applied to BPS charged black holes in four dimensions 
whose near horizon geometry is ${\rm AdS}_{2} \times {\rm S}^{2}$.
Since this near horizon geometry is similar to that of the BMPV black hole, 
the analysis of the asymptotic supersymmetry group can be extended straightforwardly 
to the four-dimensional BPS charged black holes.   

We hope to report on these problems elsewhere.

\section*{Acknowledgment}
We would like to thank T. Okuda for useful comments.
One of the authors (M. N.) also thank the Yukawa Institute for Theoretical Physics at
Kyoto University for hospitality during the workshop YITP-W-11-05 on ``Quantum Field Theory
and String Theory''. Discussions during the workshop are useful.

\appendix
\section{Lie-Lorentz derivative}
\label{sec:LL derivative}
In this appendix 
we define the Lie-Lorentz derivative \cite{Ortin:2002qb}
which is the natural extension of the standard Lie derivative.
The standard Lie derivative $\mathcal{L}$ 
does not act on any local Lorentz indices.
For example the action on the vielbein is given by 
\begin{align}
\mathcal{L}_{v}{e^{a}}_{\mu}
 =v^{\rho}\nabla_{\rho}{e^{a}}_{\mu}
 +{e^{a}}_{\rho}\nabla_{\mu}v^{\rho} 
 =-v^{\rho}{{\omega_{\rho}}^{a}}_{b}{e^{b}}_{\mu}
 +\mathcal{D}_{\mu}v^{a},
\label{eq:Lie e}
\end{align}
where $\mathcal{D}_{\mu}$ is the covariant derivative containing the both of 
the spin connection and the Christoffel symbol.
Notice that 
the term $-v^{\rho}{{\omega_{\rho}}^{a}}_{b}{e^{b}}_{\mu}$ does not transform covariantly 
since $\omega_{\mu ab}$ is the connection on the local Lorentz frame.
This means that 
the standard Lie derivative does not transform the vielbeins covariantly.
More generally, 
let us consider quantities with any local Lorentz indices.
We call such a quantity Lorentz tensor following the reference \cite{Ortin:2002qb}.
It is clear that the standard Lie derivative does not transform the Lorentz tensors covariantly.

Now we want to introduce the extended Lie derivative $\mathbb{L}$ 
which satisfies the following properties:
\begin{itemize}
\item
$\mathbb{L}$ transforms Lorentz tensors covariantly.

\item
Acting on the tensors without local Lorentz indices, 
$\mathbb{L}$ reduces to the standard Lie derivative $\mathcal{L}$.

\end{itemize}
$\mathbb{L}$ should act on a Lorentz tensor ${T_{\mu_1\cdots \mu_m}}^{\nu_1\cdots \nu_n}$ 
with the mixed curved space indices as 
\begin{align}
\mathbb{L}_{v}{T_{\mu_1\cdots \mu_m}}^{\nu_1\cdots \nu_n}
&=v^{\rho}\mathcal{D}_{\rho}{T_{\mu_1\cdots \mu_m}}^{\nu_1\cdots \nu_n}  \notag \\
&+{T_{\rho\mu_2\cdots \mu_m}}^{\nu_1\cdots \nu_n}\mathcal{D}_{\mu_1}v^{\rho}
 +\cdots   \notag \\
&-{T_{\mu_1\cdots \mu_m}}^{\rho\nu_2\cdots \nu_n}\mathcal{D}_{\rho}v^{\nu_1}
 -\cdots  \notag \\
&+\tfrac{1}{2}\epsilon^{ab}(v)\Sigma_{ab}^{(r)}{T_{\mu_1\cdots \mu_m}}^{\nu_1\cdots \nu_n},
\label{eq:general L T}
\end{align}
where $\Sigma_{ab}^{(r)}$ is a generator of Lorentz group in the representation $r$.
$\epsilon^{ab}(v)$ may be an arbitrary local Lorentz transformation parameter 
which satisfies $\epsilon^{ab}(v)=-\epsilon^{ba}(v)$.

To fix $\epsilon^{ab}(v)$ appropriately,  
let us consider the case that $v$ is a Killing vector.
Then by definition $\mathcal{L}_{v}g_{\mu\nu}=0$, 
but eq.(\ref{eq:Lie e}) reduces to
\begin{align}
\mathcal{L}_{v}{e^{a}}_{\mu}
=    -v^{\rho}{{\omega_{\rho}}^{a}}_{b}{e^{b}}_{\mu}
\neq 0.
\end{align}
It seems reasonable that we take 
\begin{align}
\mathbb{L}_{v}{e^{a}}_{\mu}=0,
\label{eq:L e vanishes}
\end{align} 
in fact this criterion reduces to
\begin{align}
0
&=\mathbb{L}_{v}{e^{c}}_{\mu}  \notag \\
&=v^{\rho}\mathcal{D}_{\rho}{e^{c}}_{\mu}
 +{e^{c}}_{\rho}\mathcal{D}_{\mu}v^{\rho}
 +\tfrac{1}{2}\epsilon^{ab}(v){\left(\Sigma_{ab}\right)^{c}}_{d}{e^{d}}_{\mu} \notag \\
&={e^{d}}_{\mu}\left(-\mathcal{D}^{c} v_{d}+{\epsilon^{c}}_{d}(v)\right)  \notag \\
&\Rightarrow
\epsilon_{ab}(v)=\mathcal{D}_{a} v_{b}.
\label{eq:form of epsilon}
\end{align}
Noting that $\mathcal{D}_{a} v_{b}=-\mathcal{D}_{b} v_{a}$, 
we find that the criterion (\ref{eq:L e vanishes}) is an appropriate one.
From eqs.(\ref{eq:general L T},\ref{eq:form of epsilon}), 
we obtain the expression
\begin{align}
\mathbb{L}_{v}{T_{\mu_1\cdots \mu_m}}^{\nu_1\cdots \nu_n}
&=v^{\rho}\mathcal{D}_{\rho}{T_{\mu_1\cdots \mu_m}}^{\nu_1\cdots \nu_n}  \notag \\
&+{T_{\rho\mu_2\cdots \mu_m}}^{\nu_1\cdots \nu_n}\mathcal{D}_{\mu_1}v^{\rho}
 +\cdots   \notag \\
&-{T_{\mu_1\cdots \mu_m}}^{\rho\nu_2\cdots \nu_n}\mathcal{D}_{\rho}v^{\nu_1}
 -\cdots  \notag \\
&+\tfrac{1}{2}\mathcal{D}^{a} v^{b}\Sigma_{ab}^{(r)}{T_{\mu_1\cdots \mu_m}}^{\nu_1\cdots \nu_n},
\label{eq:LL derivative}
\end{align}
and this is identical to the definition given in the reference \cite{Ortin:2002qb}.
Using the expression (\ref{eq:LL derivative}), 
it is showed that $\mathbb{L}$ satisfies the following properties:
\begin{itemize}
\item
The action of $\mathbb{L}_{v}$ satisfies Leibniz rule:
\begin{align}
\mathbb{L}_{v}(T_1 T_2)
=\mathbb{L}_{v}T_1 T_2
+T_1 \mathbb{L}_{v}T_2.
\end{align}

\item
$\mathbb{L}_v$ commutes with $\Gamma^a$:
\begin{align}
\left[\mathbb{L}_{v},\Gamma^a \right]T=0.
\end{align}

\item
The commutator of two Lie-Lorentz derivatives is given by
\begin{align}
\left[\mathbb{L}_{v_1},\mathbb{L}_{v_2}\right]T
=\mathbb{L}_{\left[v_1,v_2\right]_{LB}}T.
\end{align}

\item
$\mathbb{L}_{v}$ is linear in $v$.

\end{itemize}
where $T$ is a Lorentz tensor with the mixed curved space indices.

\section{Conserved charges}
\label{sec:conserved charges}
In this appendix some properties of conserved charges are reviewed.
It is convenient to start with the Lagrangian $D$-form $\bm{L}$ following the reference 
\cite{Lee:1990nz,Wald:1993nt,Iyer:1994ys,Hollands:2006zu}, 
which is related to the Lagrangian density $\mathcal{L}$ as follows:
\begin{align}
\bm{L}
=\mathcal{L}\left(d^D x\right),
\end{align}
where
$
(d^{D-p}x)_{\mu_1\cdots \mu_p}
\equiv \tfrac{1}{p!(D-p)!}\epsilon_{\mu_1\cdots \mu_p \mu_{p+1}\cdots \mu_D}
       dx^{\mu_{p+1}}\wedge\cdots \wedge dx^{\mu_D}
$
with $\epsilon_{\dot{0}\dot{1}\cdots \dot{(D-1)}}=+1$.\footnotemark
\footnotetext{
$\dot{0}\dot{1}\cdots \dot{(D-1)}$ denote the curved space indices.}
The approach which we follow here is called the covariant phase space method.

In appendix \ref{sec:conserved charge and Poisson bracket} 
we define the conserved charges and deduce some immediate consequences.
In particular, the Poisson bracket algebra of two conserved charges are discussed.
In appendix \ref{sec:construction} we construct the conserved charges 
from $D=5$ minimal supergravity action.

\subsection{Conserved charge and Poisson bracket}
\label{sec:conserved charge and Poisson bracket}
We consider the theory which is described by the Lagrangian $D$-form $\bm{L}(\Phi)$, 
where $\Phi$ denotes the dynamical fields collectively.
The variation of the Lagrangian $D$-form is given by
\begin{align}
\delta \bm{L}(\Phi)
=\bm{E}(\Phi)\delta \Phi +d\bm{\Theta}(\Phi,\delta \Phi),
\label{eq:deriving eom}
\end{align}
and the equations of motion are $\bm{E}(\Phi)=0$. 
The symmetry transformation is defined by 
\begin{align}
\delta_{\epsilon}\bm{L}(\Phi)
=d\bm{B}_{\epsilon}(\Phi).
\label{eq:sym transf}
\end{align}
The conserved charge associated with the symmetry transformation (\ref{eq:sym transf}) 
should be defined as the integration of a corresponding conserved current.
In the covariant phase space method,
such a current is defined by
\begin{align}
\bm{\omega}(\Phi,\delta_1 \Phi,\delta_2 \Phi)
\equiv \delta_1 \bm{\Theta}(\Phi,\delta_2 \Phi)
      -\delta_2 \bm{\Theta}(\Phi,\delta_1 \Phi)
      -\bm{\Theta}(\Phi,\left[\delta_1,\delta_2\right] \Phi).
\label{eq:symplectic}
\end{align}
To check that the current (\ref{eq:symplectic}) is conserved,  
we calculate $\left[\delta_1,\delta_2\right]\bm{L}$ in two different ways:
\begin{align}
\left[\delta_1,\delta_2\right]\bm{L}
=\bm{E}(\Phi)\left[\delta_1,\delta_2\right]\Phi
 +d\bm{\Theta}(\Phi,\left[\delta_1,\delta_2\right] \Phi);
\end{align}
and 
\begin{align}
\left[\delta_1,\delta_2\right]\bm{L}
&=\delta_1\left(\bm{E}(\Phi)\delta_2\Phi+d\bm{\Theta}(\Phi,\delta_2\Phi)\right)
 -(1\leftrightarrow 2)  \notag \\
&=\delta_1\bm{E}(\Phi)\delta_2\Phi-\delta_2\bm{E}(\Phi)\delta_1\Phi
 +\bm{E}(\Phi)\left[\delta_1,\delta_2\right]\Phi
 +d\left(\delta_1 \bm{\Theta}(\Phi,\delta_2\Phi)-\delta_2 \bm{\Theta}(\Phi,\delta_1\Phi)\right).
\end{align}
Then we have the conservation law 
\begin{align}
d\bm{\omega}(\Phi,\delta_1\Phi,\delta_2\Phi)
=-\delta_1\bm{E}(\Phi)\delta_2\Phi
 +\delta_2\bm{E}(\Phi)\delta_1\Phi
\approx 0.
\label{eq:omega is conserved}
\end{align}
where ``$\approx$'' denotes the onshell equality. 
Now we can define the conserved charge 
associated with the symmetry transformation (\ref{eq:sym transf}) 
by integrating the conserved current (\ref{eq:symplectic})
\begin{align}
\delta H_{\epsilon} 
\equiv \int_{C} 
       \bm{\omega}(\Phi,\delta \Phi,\delta_{\epsilon} \Phi),
\label{eq:def of charge}
\end{align}
where $C$ is a Cauchy surface.

For the existence of $H_{\epsilon}$, 
it is necessary that the definition (\ref{eq:def of charge}) satisfies the consistency condition
\begin{align}
\delta_1(\delta_2H_{\epsilon})-\delta_2(\delta_1H_{\epsilon})
=\left[\delta_1,\delta_2\right]H_{\epsilon}.
\label{eq:consistency}
\end{align}
This condition can be rewritten as
\begin{align}
0
&=
\delta_1(\delta_2H_{\epsilon})
-\delta_2(\delta_1H_{\epsilon})
-\left[\delta_1,\delta_2\right]H_{\epsilon} \notag \\
&=\int_{C}\left(
  \delta_1\bm{\omega}(\Phi,\delta_2\Phi,\delta_{\epsilon}\Phi)
 -\delta_2\bm{\omega}(\Phi,\delta_1\Phi,\delta_{\epsilon}\Phi)
 -\bm{\omega}(\Phi,\left[\delta_1,\delta_2\right]\Phi,\delta_{\epsilon}\Phi)
\right),
\label{eq:consistency 2}
\end{align}
or noting that  
$\bm{\omega}(\Phi,\delta_1\Phi,\delta_2\Phi)$ satisfies the identity
\begin{align}
0=
\delta_1\bm{\omega}(\Phi,\delta_2\Phi,\delta_3\Phi)
+\bm{\omega}(\Phi,\delta_1\Phi,\left[\delta_2,\delta_3\right]\Phi)
+(\text{cyclic terms for }\{1,2,3\}),
\end{align}
it can be rephrased as  
\begin{align}
0=
 \int_{C}
 \left(\delta_{\epsilon}\bm{\omega}(\Phi,\delta_1\Phi,\delta_2\Phi)
      +\bm{\omega}(\Phi,\delta_1\Phi,\left[\delta_2,\delta_{\epsilon}\right]\Phi)
      +\bm{\omega}(\Phi,\delta_2\Phi,\left[\delta_{\epsilon},\delta_1\right]\Phi)\right).
\label{eq:consistency 3}
\end{align}

The Poisson bracket of the two conserved charges is defined by
\begin{align}
\left[H_{\epsilon},H_{\epsilon'}\right]_{PB}
\equiv \delta_{\epsilon} H_{\epsilon'}.
\label{eq:PB}
\end{align}
To rewrite this, let us take the variation of the both sides 
\begin{align}
\delta \left[H_{\epsilon},H_{\epsilon'}\right]_{PB}
&= \delta \delta_{\epsilon} H_{\epsilon'} \notag \\
&= \delta_{\epsilon} \delta H_{\epsilon'}
  +\left[\delta,\delta_{\epsilon}\right]H_{\epsilon'}  \notag \\
&= \int_{C}\left(
   \delta_{\epsilon}\bm{\omega}(\Phi,\delta \Phi,\delta_{\epsilon'}\Phi)
  +\bm{\omega}(\Phi,\left[\delta,\delta_{\epsilon}\right] \Phi,\delta_{\epsilon'}\Phi)\right)
  \notag \\
&= \int_{C}
   \bm{\omega}(\Phi,\delta \Phi,\left[\delta_{\epsilon},\delta_{\epsilon'}\right] \Phi),
\end{align} 
where the consistency condition (\ref{eq:consistency 3}) was used 
for the last equality.
For any symmetry transformations 
we can write as 
$\left[\delta_{\epsilon},\delta_{\epsilon'}\right]\Phi=\delta_{\epsilon''}\Phi$, 
so we have
\begin{align}
&\delta \left[H_{\epsilon},H_{\epsilon'}\right]_{PB}
 = \int_{C}
   \bm{\omega}(\Phi,\delta \Phi,\delta_{\epsilon''} \Phi)
 =\delta H_{\epsilon''}, 
\end{align}
or integrating the both sides, 
\begin{align}
\left[H_{\epsilon},H_{\epsilon'}\right]_{PB}
=H_{\epsilon''}+K_{\epsilon,\epsilon'},
\end{align}
where $K_{\epsilon,\epsilon'}$ is the integral constant 
and can be interpreted as the central extension term.
Now we adjust such that $H_{\epsilon}$ vanishes for a reference field configuration 
$\Phi^{\text{ref}}$, 
then we have
\begin{align}
K_{\epsilon,\epsilon'}
 =\left[H_{\epsilon},H_{\epsilon'}\right]_{PB}
 =\int_{C}
  \bm{\omega}(\Phi,\delta_{\epsilon} \Phi,\delta_{\epsilon'}\Phi)
  |_{\Phi=\Phi^{\text{ref}}}.
\end{align}

\subsection{Construction}
\label{sec:construction}
We move on the explicit constructions of the conserved charges defined by 
eq.(\ref{eq:def of charge}).
Our first task is to rewrite the definition to the more tractable expression.
From eqs.(\ref{eq:deriving eom},\ref{eq:sym transf}) 
\begin{align}
0
=  \bm{E}(\Phi)\delta_{\epsilon} \Phi 
  +d\bm{\Theta}(\Phi,\delta_{\epsilon} \Phi)
  -d\bm{B}_{\epsilon}(\Phi).
\end{align}
Applying integration by parts to the first term
\begin{align} 
\bm{E}(\Phi)\delta_{\epsilon} \Phi
=\epsilon \bm{N}(\Phi,\bm{E}(\Phi))+d\bm{S}_{\epsilon}(\Phi,\bm{E}(\Phi)), 
\end{align}
and noting that the Noether identities imply $\bm{N}(\Phi,\bm{E}(\Phi))=0$,
then we have
\begin{align}
 d \left[\bm{S}_{\epsilon}(\Phi,\bm{E}(\Phi))
  +\bm{\Theta}(\Phi,\delta_{\epsilon} \Phi)
  -\bm{B}_{\epsilon}(\Phi) \right]
 =0,
\end{align}
or equivalently
\begin{align}
\bm{S}_{\epsilon}(\Phi,\bm{E}(\Phi))
+\bm{\Theta}(\Phi,\delta_{\epsilon} \Phi)
-\bm{B}_{\epsilon}(\Phi) 
=d\bm{Q}_{\epsilon}(\Phi).
\end{align}
Using this identity, 
we can rewrite the integrand of eq.(\ref{eq:def of charge}) as follows:
\begin{align}
&\bm{\omega}(\Phi,\delta \Phi,\delta_{\epsilon} \Phi) \notag \\
&=\delta \bm{\Theta}(\Phi,\delta_{\epsilon} \Phi)
 -\delta_{\epsilon} \bm{\Theta}(\Phi,\delta \Phi)
 -\bm{\Theta}(\Phi,\left[\delta,\delta_{\epsilon} \right] \Phi)  \notag \\
&=\delta \left(d\bm{Q}_{\epsilon}(\Phi)
               -\bm{S}_{\epsilon}(\Phi,\bm{E}(\Phi))
               +\bm{B}_{\epsilon}(\Phi)\right)
 -\delta_{\epsilon} \bm{\Theta}(\Phi,\delta \Phi)
 -\bm{\Theta}(\Phi,\left[\delta,\delta_{\epsilon} \right] \Phi) \notag \\
&\approx
 d\delta \bm{Q}_{\epsilon}(\Phi)
 +\delta \bm{B}_{\epsilon}(\Phi)
 -\delta_{\epsilon} \bm{\Theta}(\Phi,\delta \Phi)
 -\bm{\Theta}(\Phi,\left[\delta,\delta_{\epsilon} \right] \Phi),
\end{align}
where $\delta \bm{S}_{\epsilon}(\Phi,\bm{E}(\Phi)) \approx 0$ was used in the last line.
From the conservation law (\ref{eq:omega is conserved})
\begin{align}
-\left(
\delta \bm{B}_{\epsilon}(\Phi)
-\delta_{\epsilon} \bm{\Theta}(\Phi,\delta \Phi)
-\bm{\Theta}(\Phi,\left[\delta,\delta_{\epsilon} \right] \Phi)
\right)
\approx 
 d\delta \bm{Q}_{\epsilon}(\Phi)
-\bm{\omega}(\Phi,\delta \Phi,\delta_{\epsilon} \Phi)
\approx 
d\bm{A}_{\epsilon}(\Phi,\delta \Phi),
\end{align}
so we have
\begin{align}
\bm{\omega}(\Phi,\delta \Phi,\delta_{\epsilon} \Phi)
\approx d\bm{k}_{\epsilon}(\Phi,\delta \Phi),  \quad
\bm{k}_{\epsilon}(\Phi,\delta \Phi)
\equiv \delta \bm{Q}_{\epsilon}(\Phi)-\bm{A}_{\epsilon}(\Phi,\delta \Phi),
\end{align}
and this means that the eq.(\ref{eq:def of charge}) reduces to
\begin{align}
\delta H_{\epsilon}\left[\Phi \right]
\approx \int_{\partial C}\bm{k}_{\epsilon}(\Phi,\delta \Phi).
\end{align}

\paragraph{Bosonic symmetry}
Let us apply the algorithm described above to $D=5$ minimal supergravity 
whose action is given by eq.(\ref{eq:action}).
Here we focus on the bosonic symmetries 
which consist of general coodinate transformations and $U(1)$ gauge transformations.
For all dynamical fields 
the general coodinate transformation is represented by
$\delta_{v}\Phi=\mathbb{L}_{v}\Phi$, 
and the $U(1)$ gauge transformation acts on the only gauge field $A_{\mu}$ as 
$\delta_{\Lambda}A_{\mu}=\nabla_{\mu}\Lambda$.

Notice that there are no contributions from the action with more than one gravitino fields, 
because we are interested in the background where the gravitino vanishes.
Therefore we consider the contributions from 
the Einstein-Hilbert term $\bm{L}_{E}$, 
the Maxwell term $\bm{L}_{F}$ 
and the Chern-Simons term $\bm{L}_{CS}$ only.

First we consider the Chern-Simons contributions.
The Lagrangian $5$-form and the symmetry transformation are given by
\begin{align}
\bm{L}_{CS}
=\tfrac{1}{6\sqrt{3}\pi}A\wedge F\wedge F, \qquad
\delta_{v,\Lambda} A
=d(v\cdot A +\Lambda)+v\cdot F,
\end{align}
respectively, 
so we have
\begin{align}
&\bm{\Theta}^{CS}(A,\delta A)
=-\tfrac{1}{3\sqrt{3}\pi}A\wedge F\wedge \delta A, \qquad
\bm{S}^{CS}_{v,\Lambda}
=\tfrac{1}{2\sqrt{3}\pi}F\wedge F (v\cdot A +\Lambda),   \\
&\bm{B}^{CS}_{v,\Lambda}(A)
= \tfrac{1}{6\sqrt{3}\pi} \left[
  v\cdot (A\wedge F\wedge F) +\Lambda F\wedge F  \right],
\end{align}
and
\begin{align}
\bm{A}^{CS}_{v,\Lambda}(A,\delta A)
&=\tfrac{1}{3\sqrt{3}\pi}\left(
  -\delta A\wedge F (v\cdot A+\Lambda)
  +A\wedge (v\cdot F)\wedge \delta A
  +A\wedge F(v \cdot \delta A)\right),   \\
\bm{Q}^{CS}_{v,\Lambda}(A)
&=\tfrac{1}{3\sqrt{3}\pi}A\wedge F(v\cdot A +\Lambda).
\end{align}
Therefore the contribution from $\bm{L}_{CS}$ term is given by
\begin{align}
\bm{k}^{CS}_{v,\Lambda}(A,\delta A)
&=\tfrac{1}{\sqrt{3}\pi}\delta A\wedge F(v\cdot A +\Lambda)
  +\tfrac{1}{3\sqrt{3} \pi}A\wedge \delta A\wedge  \delta _{v,\Lambda}A
  +d\left(\tfrac{1}{3\sqrt{3}\pi}\delta A\wedge A(v\cdot A +\Lambda)\right).
\label{eq:superpotential CS}
\end{align}
Note that the last term of eq.(\ref{eq:superpotential CS}) does not contribute conserved charges, 
since conserved charges are given by the integration of eq.(\ref{eq:superpotential CS}) 
on the boundary of a Cauchy surface.

Next we consider the Einstein-Hilbert and Maxwell contributions. 
Noting that 
\begin{align}
\delta_{\Lambda} \bm{L}_{E/F}=0,
\end{align} 
$\bm{B}_{v,\Lambda}(\Phi)$ and $\bm{A}_{v,\Lambda}(\Phi,\delta \Phi)$ are written as
\begin{align}
\bm{B}^{E/F}_{v,\Lambda}(\Phi)=v\cdot \bm{L}, \qquad
\bm{A}^{E/F}_{v,\Lambda}(\Phi,\delta \Phi) =v\cdot \bm{\Theta}(\Phi,\delta \Phi).
\end{align}
Therefore the expression of the conserved charge reduces to
\begin{align}
\delta H^{E/F}_{v,\Lambda}\left[\Phi\right]
\approx \int_{\partial C}\bm{k}^{E/F}_{v,\Lambda}(\Phi,\delta \Phi), \qquad
\bm{k}^{E/F}_{v,\Lambda}(\Phi,\delta \Phi)
\equiv \delta \bm{Q}^{E/F}_{v,\Lambda}(\Phi)-v\cdot \bm{\Theta}^{E/F}(\Phi,\delta \Phi),
\end{align} 
where
\begin{align}
d\bm{Q}^{E/F}_{v,\Lambda}(\Phi)
=
\bm{S}^{E/F}_{v,\Lambda}(\Phi,\bm{E}(\Phi))
+\bm{\Theta}^{E/F}(\Phi,\delta_{v,\Lambda} \Phi)
-v\cdot \bm{L}_{E/F}.
\end{align}
The Einstein-Hilbert Lagrangian $D$-form is given by
\begin{align}
\bm{L}_{E}
=\tfrac{1}{16 \pi}\sqrt{-g}R \left(d^D x\right)
\end{align}
so we have
\begin{align}
\bm{\Theta}^{E}(\Phi,\delta \Phi)
=\tfrac{\sqrt{-g}}{16\pi}\left(\nabla_{\nu}h^{\mu\nu}-\nabla^{\mu}h\right)
   \left(d^{D-1} x\right)_{\mu},  \qquad
\bm{S}^{E}_{v}
=-\tfrac{\sqrt{-g}}{8\pi}G^{\mu\nu}v_{\nu}
   \left(d^{D-1} x\right)_{\mu},
\end{align}
and 
\begin{align}
\bm{Q}^{E}_{v,\Lambda}(\Phi)
=\tfrac{\sqrt{-g}}{8\pi}\nabla^{\nu}v^{\mu}
  \left(d^{D-2} x\right)_{\mu\nu}.
\end{align}
Therefore the contribution from $\bm{L}_{E}$ term is given by
\begin{align}
\bm{k}^{E}_{v}(\Phi,\delta \Phi)
&=\frac{\sqrt{-g}}{8\pi}\left[
  v^{\nu}\nabla^{\mu}h
 -v^{\nu}\nabla_{\rho}h^{\mu\rho}
 +v_{\sigma}\nabla^{\nu}h^{\mu\sigma}
 +\frac{1}{2}h\nabla^{\nu}v^{\mu}
 -h^{\nu\rho}\nabla_{\rho}v^{\mu}\right]
 \left(d^{D-2} x\right)_{\mu\nu}.
\end{align}
Similarly the contribution from $\bm{L}_{F}$ term is given by
\begin{align}
\bm{k}^{F}_{v,\Lambda} (\Phi,\delta \Phi)
&=\frac{\sqrt{-g}}{16\pi}\left[
  \left(-2hF^{\mu\nu}+8h^{\rho\mu}{F_{\rho}}^{\nu}-8\nabla^{\mu}a^{\nu}\right)
  (A_{\rho}v^{\rho}+\Lambda)  \right.  \notag \\
& \left.\hspace{4cm} 
 -4F^{\mu\nu}a_{\rho}v^{\rho}
 -8F^{\nu\rho}a_{\rho}v^{\mu}\right]
 \left(d^{D-2} x\right)_{\mu\nu}.
\end{align}

\paragraph{Supersymmetry}
Finally, we derive the conserved charges associated with the supersymmetry transformations
(\ref{eq:susy e}-\ref{eq:susy A}).
Noting that we remain on the background where the gravitino vanishes, 
it is showed that the only contribution comes from $\bm{L}_{\psi}$ term 
which is given by
\begin{align}
\bm{L}_{\psi}
=-\tfrac{i}{8 \pi}e(\overline{\psi}_{\mu}\Gamma^{\mu\nu\rho}D_{\nu}\psi_{\rho}
     +\overline{\psi}_{\rho}\overleftarrow{D}_{\nu}\Gamma^{\mu\nu\rho}\psi_{\mu})
 \left(d^5 x\right).
\end{align}
Then we have
\begin{align}
\bm{\Theta}^{\psi}(\Phi,\delta \Phi)
=\tfrac{e}{16 \pi}\left[-2i \overline{\delta \psi_{\nu}}\Gamma^{\mu\nu\rho}\psi_{\rho}
 +\text{(h.c.)}
 +\mathcal{O}(\psi^2)\right] \left(d^4 x\right)_{\mu}.
\label{eq:Theta psi susy}
\end{align}
In this case 
it is easy task to construct the conserved charge 
from the definition (\ref{eq:def of charge}) directly, 
since we are not interested in 
the explicit forms of the higher order terms with respect to gravitinos.
The only term which is relevant to our calculation 
is the $\mathcal{O}(\psi^0)$ term in the integrand of the conseved charge.
From eq.(\ref{eq:Theta psi susy}) we have
\begin{align}
\bm{\omega}^{\psi}(\Phi,\delta \Phi, \delta_{\xi} \Phi)
&=\delta \bm{\Theta}^{\psi}(\Phi,\delta_{\xi} \Phi)
 -\delta_{\xi} \bm{\Theta}^{\psi}(\Phi,\delta \Phi)
 -\bm{\Theta}^{\psi}(\Phi,\left[\delta,\delta_{\xi} \right] \Phi)  \notag \\
&=\tfrac{1}{4\pi}\left[ 
  e\nabla_{\rho}\left(i\overline{\xi}\Gamma^{\mu\nu\rho}\delta \psi_{\nu}\right) \right. \notag \\
& \left. \hspace{2cm}
 -\tfrac{1}{4}e\overline{\xi}\left(-4i\Gamma^{\mu\nu\rho}D_{\nu}\delta \psi_{\rho}
                                 +\sqrt{3}X^{\mu\nu\rho\sigma}\delta \psi_{\nu}F_{\rho\sigma}\right)
 +\text{(h.c.)}
 +\mathcal{O}(\psi) \right] \left(d^4 x\right)_{\mu}  \notag \\
&\approx
 e\nabla_{\nu} 
 \left(-\tfrac{i}{4\pi} 
 \overline{\xi}\Gamma^{\mu\nu\rho}\delta \psi_{\rho} 
 +\text{(h.c.)}
 +\mathcal{O}(\psi) \right)\left(d^4 x\right)_{\mu}  \notag \\
&=d\bm{k}^{\psi}_{\xi}(\Phi,\delta \Phi),
\end{align}
where the linearized equations of motion was used in the third line, and
\begin{align}
\bm{k}^{\psi}_{\xi}(\Phi,\delta \Phi)
&=e\left[-\tfrac{i}{4\pi}\overline{\xi} \Gamma^{\mu\nu\rho} \delta \psi_{\rho}
 +\text{(h.c.)}
 +\mathcal{O}(\psi) \right] \left(d^3 x\right)_{\mu\nu}.
\label{eq:superpotential susy}
\end{align}
The last term of eq.(\ref{eq:superpotential susy}) vanishes 
when we are interested in the $\psi_{\mu}=0$ background.

\end{document}